\documentclass[aps,physrev,pra,preprint,groupedaddress,unsortedaddress,amsmath,amssymb]{revtex4-2}

\usepackage{graphicx}
\usepackage{dcolumn}
\usepackage{bm}
\usepackage{siunitx}
\usepackage{mathtools}
\usepackage{enumitem}
\usepackage{hyperref} 
\hypersetup{
    colorlinks=true,       
    citecolor=blue,      
	linkcolor=black,      
    urlcolor=black,        
}
\begin{document}


\title{\textbf{Wave-optical formulation of the image-rotation property in Dove prisms: A Fourier-optics approach}}
\author{Nima Keramati}
\affiliation{%
 Department of Electrical Engineering, Semnan University, Semnan 3513119111, Iran
}%
\author{S. Faezeh Mousavi}
 
\affiliation{%
 Department of Physics, University of Trieste, Trieste 34127, Italy
}%
\affiliation{%
 National Institute of Optics (CNR-INO), Basovizza 34149, Italy
}%

\author{Amirnader Askarpour}
\affiliation{
 Amirkabir University of Technology, Tehran 1591634311, Iran
}%

\author{Majid Afsahi}
\email{m\_afsahi@semnan.ac.ir}
 \affiliation{%
 Department of Electrical Engineering, Semnan University, Semnan 3513119111, Iran
}%




\begin{abstract}
In this paper, we present a formula for calculating the complex amplitude of the output electric field for a given input wave that impinges on a Dove prism. We use Fourier optics to decompose the input wave into plane waves, then find the output plane waves of the Dove prism as functions of the input spatial frequencies. The total output image is then obtained by integrating over all the output plane waves, resulting in a final formula in integral form. Since we conduct a wave-optical analysis for beam propagation and each incidence on Dove prism surfaces, all the physical aspects of electromagnetic waves are involved, including polarization, Fresnel losses, wave interference, phase, and intensity. The formula also explains why a rotated Dove prism rotates its input image twice its rotation angle. In addition, the formula is not limited to paraxial beams, as we find the Dove prism output as a function of the input Fourier components in general, without limiting the input spatial frequencies to small values. This generality is especially relevant for emerging applications that rely on non-paraxial beams, such as structured light generation, orbital angular momentum systems, and quantum imaging. However, since in most cases the paraxial approximation is valid and sufficient, a simplified formula is also extracted for paraxial beams. Two ray-tracing simulations are conducted to demonstrate the correctness and accuracy of our simplified formula. All the advantages mentioned make our derivation accurate, complete, comprehensive, and, to the best of our knowledge, the first to wave-optically prove the rotational feature of a Dove prism.      
\end{abstract}

\maketitle
\section{\label{sec:intro}Introduction}
A Dove prism is an image-rotating optical component and is highly practical in setups in which image inversion or rotation is needed. It has applications in areas such as rotary shearing interferometry \cite{armitageRotaryShearingInterferometry1965}, optical orbital angular momentum (OAM) sorting \cite{leachInterferometricMethodsMeasure2004}, and surface plasmon resonance spectroscopy \cite{bolducHighresolutionSurfacePlasmon2009}. In the scope of its image rotation application, the well-known performance of a Dove prism is that if it is rotated through $\alpha$ around its longitudinal axis, its input image inverts vertically and undergoes a rotation of $2\alpha$. This principle has been successfully exploited in numerous experiments and incorporated into the design of optical systems. Nevertheless, apart from the physical tests, it has generally been treated as a postulate; since there is no complete mathematical proof for the Dove prism rotation effect. In fact, to the best of our knowledge, only a geometrical-optical justification for Dove image rotation was given in \cite{smithModernOpticalEngineering2008}. There have also been studies on other aspects of the functionality or structure of a Dove prism such as geometrical-optical analysis of the ellipticity introduced by a Dove prism \cite{gonzalezHowDovePrism2006}, polarization properties of a Dove prism \cite{padgettDovePrismsPolarized1999,morenoPolarizationTransformingProperties2003,morenoJonesMatrixImagerotation2004,karanQuantifyingPolarizationChanges2022,wangSinglepathSagnacInterferometer2017}, modification of the formulae related to the length and weight of a Dove prism \cite{sar-elRevisedDovePrism1991}, and formulation of the extent of Dove prism wave front tilt caused by manufacturing errors \cite{morenoDovePrismIncreased2003}.     

Thus, there are two gaps in the research areas related to the Dove prism. First, all existing system designs, calculations, predictions, and experiments, although correct, have been based on an unproven experimental physical property. Second, there is a lack of a formula to exactly determine the complex electric field at the output of a Dove prism, particularly when dealing with non-paraxial beams, for which a simple image rotation modeling is insufficient. There are a variety of applications that require such knowledge of a Dove prism, including high-precision interferometers \cite{versmold_interferometric_2025,zhang_hong-ou-mandel_2016}, sensitivity analysis and Dove prism vibration in optical setups \cite{xiao_orbital_2018,yao_loss-tolerant_2024}, accurate OAM sorting for quantum information applications \cite{wangSinglepathSagnacInterferometer2017}, astronomical applications \cite{montes-flores_rotationally_2024}, complex field generation \cite{chen_precise_2017}, and imaging systems \cite{lowry_reflection_2021,byersSuperresolutionUpgradeDeep2025}. By knowing an accurate formula for the Dove prism, we can avoid a computationally costly direct numerical solution of Maxwell's equations if an exact electric field is needed at the output. Additionally, having a pre-existing formula is helpful whenever optimization is required. It is worth mentioning that ray-tracing or geometrical optics techniques available in many commercial software packages, although fast, fail to produce correct and consistent results in the non-paraxial regime, such as when beam divergence or high spatial frequencies are involved.

In this paper, using Fourier analysis, a comprehensive mathematical proof is given to justify the image rotation property of a Dove prism with a $\SI{45}{\degree}$ base angle. We derived a general formula (Eq. \eqref{eq:uo}) from which both phase and amplitude patterns of the output image can be obtained. Our formula elucidates all the Dove prism effects, including image inversion and rotation, polarization changes, and Fresnel losses. It is not restricted to small spatial frequencies; therefore, it can cover applications as was done in \cite{montes-flores_rotationally_2024}. All of these effects are integrated into a single closed-form integral. For verification, we performed two ray-tracing simulations and compared the results with those obtained using our final formula. 
\section{\label{sec:analysis}Fourier Analysis and Analytical Derivation of the Dove Prism Image Rotating Feature}
\begin{figure}
\includegraphics{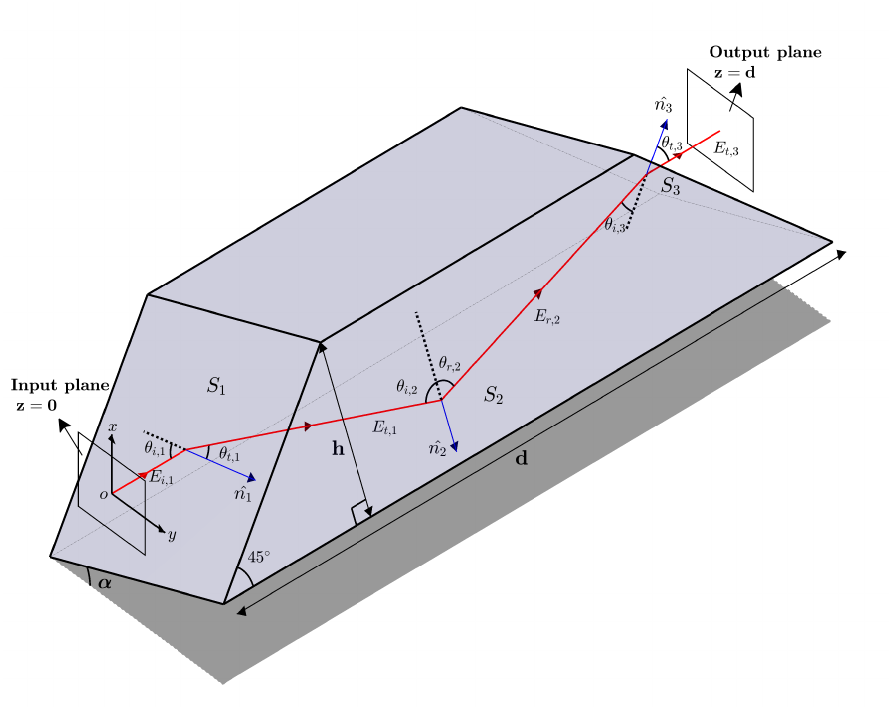}
\caption{\label{fig:dove}Schematic of a Rotated Dove prism. The prism is rotated through $\alpha$ around the $z$-axis, and its length and height are $d$ and $h$, respectively. The red lines represent the trajectory of the beam from the input to the output plane. The blue vectors represent the normal unit vectors of the surfaces, and the dotted lines are their continuation towards the opposite direction.} 
\end{figure}

Assume that the input beam of the Dove prism propagates along the $z$ direction and has an electric field complex phasor ${{\vec{E}}_{in}}={{\hat{a}}_{in}}U_{in}(x,y,z)$, in which $U_{in}(x,y,z)$ is the complex amplitude of the input beam, and  $\hat{a}_{in}$ is the unit polarization vector. The approach that we adopted to obtain the output electric field of a Dove prism is based on Fourier optics. The idea is to decompose the input beam into its constituting plane waves and to find their corresponding outputs, which are also plane waves. Once we have the output waves, the final image can be obtained simply by integrating over all of the output plane waves. To perform this idea, we consider a general Fourier component with an electric field phasor ${{\vec{E}}_{F_{in}}}={{\hat{a}}_{in}}\left[F_{in}\left({{k}_{in_x}},{{k}_{in_{y}}}\right)\exp\boldsymbol{(}-j({{k}_{in_{x}}}x+{{k}_{in_{y}}}y+{{k}_{in_{z}}}z)\boldsymbol{)}\right]$, in which $k_{in_x}$, $k_{in_y}$, and $k_{in_z}$ are the components of the wave vector $\vec{k}_{in}$, and $F_{in}(k_{in_x},k_{in_y})$ is the two-dimensional Fourier transform of $U_{in}$ at the input plane of the Dove prism ($z=0$), which is given by the following equation:
\begin{equation}
F_{in}({{k}_{in_x}},{{k}_{in_y}})=\int\limits_{-\infty }^{\infty }{\int\limits_{-\infty }^{\infty }{U_{in}(x,y,0)\exp \boldsymbol{(}j({{k}_{in_x}}x+{{k}_{in_y}}y)\boldsymbol{)}dxdy}}.
    \label{eq1}
\end{equation}
The polarization of all the Fourier components is the same as the original beam, while their propagation direction varies with $k_{in_x}$ and $k_{in_y}$. The effect of a rotated Dove prism on ${{\vec{E}}_{F_{in}}}$  is schematically represented in Fig. \ref{fig:dove}. The incident plane wave that emerges from the $z=0$ plane undergoes a refraction at the surface $S_1$, a total internal reflection at $S_2$, and a second refraction at $S_3$. The wave exiting $S_3$ is the output wave that hits the output plane at $z=d$ and has the electric phasor ${{\vec{E}}_{F_o}}$ and wave vector $\vec{k}_o=k_{o_x}\hat{x}+k_{o_y}\hat{y}+k_{o_z}\hat{z}$. The center of the input plane is set as the origin of the global coordinate system $(x, y, z)$. Deriving the output for a single input Fourier component is straightforward. We use the equality of the tangential components of electric and magnetic fields at each plane of incidence. To do so, it is first required that the incident wave be decomposed into TE and TM polarizations. The equalities of tangential components must then be formed independently for each polarization, yielding two pairs of equations. Solving these two equation systems gives us the TE and TM parts of the complex amplitude of the output wave—the transmitted wave at $S_1$ and $S_3$ and the reflected wave at $S_2$—and a relation between the incident and output spatial frequencies. Thus, for each surface , the total output amplitude and net polarization are obtained by summing the TE and TM parts, and the direction of propagation is given by the output spatial frequencies, which are obtained from the spatial frequencies of the incident wave. The output of each surface will be the incident wave for the next one, and this procedure continues until ${{\vec{E}}_{F_o}}$ is found.

To express the formulae comprehensively, it is required to define a general form for three base vectors (${{\hat{x}}_{l}},{{\hat{y}}_{l}},{{\hat{z}}_{l}}$) for the surface $l$ of the Dove prism. By considering ${{\vec{k}}_{i,l}}={{k}_{i,l_x}}\hat{x}+{{k}_{i,l_y}}\hat{y}+{{k}_{i,l_z}}\hat{z}$ as the wave vector of the incident wave at $S_l$, and by using the same notation as \cite{chengFieldWaveElectromagnetics1989}, we choose ${{\hat{z}}_{l}}={{\hat{n}}_{l}}$, ${{\hat{y}}_{l}}=\frac{{{\hat{n}}_{l}}\times {{{\vec{k}}}_{i,l}}}{\left| {{{\hat{n}}}_{l}}\times {{{\vec{k}}}_{i,l}} \right|}$, and ${{\hat{x}}_{l}}={{\hat{y}}_{l}}\times {{\hat{z}}_{l}}$, where ${{\hat{n}}_{l}}$ is the normal unit vector of $S_l$, as depicted in Fig. \ref{fig:dove}. The TE polarization unit vector will then be ${{\hat{s}}_{l}}={{\hat{y}}_{l}}$, and the TM polarization unit vectors are defined as ${{\hat{p}}_{i,l}}=\cos {{\theta }_{i,l}}{{\hat{x}}_{l}}-\sin {{\theta }_{i,l}}{{\hat{z}}_{l}}$ for the incident wave, ${{\hat{p}}_{r,l}}=\cos {{\theta }_{r,l}}{{\hat{x}}_{l}}+\sin {{\theta }_{r,l}}{{\hat{z}}_{l}}$ for the reflected wave, and ${{\hat{p}}_{t,l}}=\cos{{\theta }_{t,l}}{{\hat{x}}_{l}}-\sin {{\theta }_{t,l}}{{\hat{z}}_{l}}$ for the transmitted wave, where ${{\theta }_{i,l}}=\arcsin\left( \frac{{{{\vec{k}}}_{i,l}}\cdot {{{\hat{x}}}_{l}}}{\left| {{{\vec{k}}}_{i,l}} \right|} \right)$, ${{\theta }_{r,l}}={{\theta }_{i,l}}$, and ${{\theta }_{t,l}}=\arcsin\left(\frac{\eta_{t,l}}{\eta_{i,l}}\sin {{\theta }_{i,l}}\right)$ are, respectively, the incident, reflected, and refracted angles, with $\eta_{i,l}$ and $\eta_{t,l}$ the intrinsic impedance of the incident and transmitted media. In general, the conversion of the polarization components of an electric field from its previous base vectors $({{\hat{s}}_{pre}},{{\hat{p}}_{pre}})$ to the new basis $({{\hat{s}}_{new}},{{\hat{p}}_{new}})$ is done by the transformation matrix $\mathbf{{T}_{(pre),(new)}}=\begin{bsmallmatrix}
   {{{\hat{s}}}_{new}}.{{{\hat{s}}}_{pre}} & {{{\hat{s}}}_{new}}.{{{\hat{p}}}_{pre}}  \\
   {{{\hat{p}}}_{new}}.{{{\hat{s}}}_{pre}} & {{{\hat{p}}}_{new}}.{{{\hat{p}}}_{pre}}  \\
\end{bsmallmatrix}$.

From the above definitions, one can write the combined TE and TM electric fields of the incident, reflected, and refracted waves at $S_l$ as follows:
\begin{subequations}
    \begin{equation}    \vec{E}_{i,l}=\left[E_{{i,l}_{s}}\hat{s}_l+E_{i,l_p}\hat{p}_{i,l}\right]\exp\boldsymbol{(}-j(k_{i,l_x}x+k_{i,l_y}y+k_{i,l_z}z)\boldsymbol{)},
    \label{eq:incident_general}
    \end{equation}
    \begin{equation}    \vec{E}_{r,l}=\left[E_{r,l_s}\hat{s}_l+E_{r,l_p}\hat{p}_{r,l}\right]\exp\boldsymbol{(}-j(k_{r,l_x}x+k_{r,l_y}y+k_{r,l_z}z)\boldsymbol{)},
    \label{eq:reflection_general}
    \end{equation}    
    \begin{equation}    \vec{E}_{t,l}=\left[E_{t,l_s}\hat{s}_l+E_{t,l_p}\hat{p}_{t,l}\right]\exp\boldsymbol{(}-j(k_{t,l_x}x+k_{t,l_y}y+k_{t,l_z}z)\boldsymbol{)}.
    \label{eq:transmitted_general}
    \end{equation}
    \label{eq:fields_general}
\end{subequations}\\
The subscripts $s$ and $p$ denote TE and TM polarizations, respectively. All magnetic fields can be found from their corresponding electric fields using the relation $\vec{H}=\frac{1}{\eta}\frac{\vec{k}}{|\vec{k}|}\times\vec{E}$. Note that in the exponential terms of Eq. \eqref{eq:fields_general} we use the global coordinate system $(x,y,z)$.
 
The critical step for finding the outputs on each prism surface is to form the equations that arise from the continuity of the tangential electric and magnetic fields, which are always aligned with $\hat{x}_l$ and $\hat{y}_l$. In doing so, for the surface $S_l$ we obtain:

\begin{subequations}
    \begin{equation}
    \begin{split}
       &E_{i,l_s}\exp\boldsymbol{(}-j(k_{i,l_x}x+k_{i,l_y}y+k_{i,l_z}z)\boldsymbol{)}\\&+E_{r,l_s}\exp\boldsymbol{(}-j(k_{r,l_x}x+k_{r,l_y}y+k_{r,l_z}z)\boldsymbol{)}\\&=E_{t,l_s}\exp\boldsymbol{(}-j(k_{t,l_x}x+k_{t,l_y}y+k_{t,l_z}z)\boldsymbol{)},
       \end{split}
    \end{equation}
    \\
    \begin{equation}
    \begin{split}
        &-\frac{\cos(\theta_{i,l})}{\eta_{i,l}}E_{i,l_s}\exp\boldsymbol{(}-j(k_{i,l_x}x+k_{i,l_y}y+k_{i,l_z}z)\boldsymbol{)}\\&+\frac{\cos(\theta_{r,l})}{\eta_{i,l}}E_{r,l_s}\exp\boldsymbol{(}-j(k_{r,l_x}x+k_{r,l_y}y+k_{r,l_z}z)\boldsymbol{)}\\&=-\frac{\cos(\theta_{t,l})}{\eta_{t,l}}E_{t,l_s}\exp\boldsymbol{(}-j(k_{t,l_x}x+k_{t,l_y}y+k_{t,l_z}z)\boldsymbol{)},
        \end{split}
    \end{equation}
    \label{eq:tangentialte_eqs}
\end{subequations}\\
for TE polarization, and 

\begin{subequations}
    \begin{equation}
    \begin{split}
      &\cos(\theta_{i,l})E_{i,l_p}\exp\boldsymbol{(}-j(k_{i,l_x}x+k_{i,l_y}y+k_{i,l_z}z)\boldsymbol{)}\\&+\cos(\theta_{r,l})E_{r,l_p}\exp\boldsymbol{(}-j(k_{r,l_x}x+k_{r,l_y}y+k_{r,l_z}z)\boldsymbol{)}
       \\&=\cos(\theta_{t,l})E_{t,l_p}\exp\boldsymbol{(}-j(k_{t,l_x}x+k_{t,l_y}y+k_{t,l_z}z)\boldsymbol{)},  
    \end{split}        
    \end{equation}
    \\
    \begin{equation}
    \begin{split}
        &\frac{1}{\eta_{i,l}}E_{i,l_p}\exp\boldsymbol{(}-j(k_{i,l_x}x+k_{i,l_y}y+k_{i,l_z}z)\boldsymbol{)}\\&-\frac{1}{\eta_{i,l}}E_{r,l_p}\exp\boldsymbol{(}-j(k_{r,l_x}x+k_{r,l_y}y+k_{r,l_z}z)\boldsymbol{)}\\&=\frac{1}{\eta_{t,l}}E_{t,l_p}\exp\boldsymbol{(}-j(k_{t,l_x}x+k_{t,l_y}y+k_{t,l_z}z)\boldsymbol{)}.
        \end{split}
    \end{equation}
    \label{eq:tangentialtm_eqs}
\end{subequations}\\
for TM polarization.

The input Fourier component ${\vec{E}}_{F_{in}}$ impinges on $S_1$ as the incident field ${\vec{E}}_{i,1}$. Assume that the input polarization ${{\hat{a}}_{in}}$ can be expressed by the Jones matrix $\begin{bsmallmatrix}
   {{{\hat{a}}}_{in}}.{{{\hat{s}}}_{in}}  \\
   {{{\hat{a}}}_{in}}.{{{\hat{p}}}_{in}}  \\
\end{bsmallmatrix}$, in which ${{\hat{s}}_{in}}=\hat{y}$, and ${{\hat{p}}_{in}}=\frac{\hat{y}\times \vec{k}_{in}}{\left| \hat{y}\times \vec{k}_{in} \right|}$. We then have $\vec{E}_{F_{in}}=[E_{F_{in_y}}\hat{s}_{in}+E_{F_{in_{xz}}}\hat{p}_{in}]\exp(-j\vec{k}_{in}\cdot\vec{r})$, where $\vec{r}=\hat{x}+\hat{y}+\hat{z}$. Thus, $\vec{E}_{i,1}=[E_{i,1_s}\hat{s}_{1}+E_{i,1_p}\hat{p}_{i,1}]\exp(-j\vec{k}_{in}\cdot\vec{r})$, where 
\begin{equation}
    \left[ \begin{matrix}
   {{E}_{i,1_s}}  \\
   {{E}_{i,1_p}}  \\
\end{matrix} \right]=\left[ \begin{matrix}
   {{{\hat{s}}}_{1}}.{{{\hat{s}}}_{in}} & {{{\hat{s}}}_{1}}.{{{\hat{p}}}_{in}}  \\
   {{{\hat{p}}}_{i,1}}.{{{\hat{s}}}_{in}} & {{{\hat{p}}}_{i,1}}.{{{\hat{p}}}_{in}}  \\
\end{matrix} \right]\left[ \begin{matrix}
   {{E}_{F_{in_y}}}  \\
   {{E}_{F_{in_{xz}}}}  \\
\end{matrix} \right]=\mathbf{{T}_{(F_{in}),(i,1)}}\left[ \begin{matrix}
   {{E}_{F_{in_y}}}  \\
   {{E}_{F_{in_{xz}}}}  \\
\end{matrix} \right].
\label{eq:einc1}
\end{equation}
Note that the propagation direction of ${{\vec{E}}_{i,1}}$ is the same as ${{\vec{E}}_{F_{in}}}$, that is ${\vec{k}}_{i,1}={\vec{k}}_{in}$.  If we choose the rotation angle $\alpha $ as the angle from the positive $y$-axis to positive $x$-axis, as depicted in Fig. \ref{fig:dove}, the plane equation of $S_1$ will be $z=\frac{h}{2}+\cos \left( \alpha  \right)x-\sin \left( \alpha  \right)y$. Putting this into the continuity equations Eq. \eqref{eq:tangentialte_eqs} and Eq. \eqref{eq:tangentialtm_eqs}, and by eliminating ${\vec{E}}_{r,1}$, we will find $\vec{E}_{t,1}=[E_{t,1_s}\hat{s}_{1}+E_{t,1_p}\hat{p}_{t,1}]\exp(-j\vec{k}_{t,1}\cdot\vec{r})$ as follows:
\begin{subequations}
    
\begin{equation}
    \left[ \begin{matrix}
   {{E}_{t,1_s}}  \\
   {{E}_{t,1_p}}  \\
\end{matrix} \right]=H_1\boldsymbol{\tau_{1}}\left[ \begin{matrix}
   {{E}_{i,1_s}}  \\
   {{E}_{i,1_p}}  \\
\end{matrix}\right],
\label{eq:et1_components}
\end{equation}

\begin{equation}
 {{k}_{t,1_x}}=\cos (\alpha )({{k}_{in_z}}-{{k}_{t,1_z}})+{{k}_{in_x}},
 \label{eq:et1_kxt1}
\end{equation}
\begin{equation}
    {{k}_{t,1_y}}=-\sin (\alpha )({{k}_{in_z}}-{{k}_{t,1_z}})+{{k}_{in_y}},
    \label{eq:et1_kyt1}
\end{equation}
\label{eq:et1}
\end{subequations}\\
where ${{H}_{1}}=\exp \boldsymbol{(}-j({{k}_{in_z}}\frac{h}{2}-{{k}_{t,1_z}}\frac{h}{2})\boldsymbol{)}$, and $\boldsymbol{\tau_{1}}=\begin{bsmallmatrix}
   {{\tau }_{1_{s}}} & 0  \\
   0 & {{\tau }_{1_{p}}}  \\
\end{bsmallmatrix}$ is the Fresnel transmission coefficient matrix for the incidence at $S_1$ (The formulas of Fresnel coefficients are available in appendix \ref{sec:appendixA}). Notice that a dispersion relation ${{k}_{t,1_z}}=\sqrt{{{({{n}}k_{in})}^{2}}-k_{t,1_x}^{2}-k_{t,1_y}^{2}}$ is also held for the transmitted wave, where ${{n}}$ is the refractive index of glass (or, in general, the ratio of the inside index to the outside index), and $k_{in}=\frac{2\pi}{\lambda}$ is the wave number of the input beam having a wavelength of $\lambda$.

 ${{\vec{E}}_{t,1}}$ is considered as the incident wave at $S_2$, ${{\vec{E}}_{i,2}}$, and will totally reflect towards $S_3$ as ${{\vec{E}}_{r,2}}$. We follow the same procedure to find ${{\vec{E}}_{r,2}}$ by first rewriting ${{\vec{E}}_{t,1}}$ as the incident wave $\vec{E}_{i,2}=[E_{i,2_s}\hat{s}_{2}+E_{i,2_p}\hat{p}_{i,2}]\exp(-j\vec{k}_{t,1}\cdot\vec{r})$,  where
\begin{equation}
    \left[ \begin{matrix}
   {{E}_{i,2_s}}  \\
   {{E}_{i,2_p}}  \\
\end{matrix} \right]=\left[ \begin{matrix}
   {{{\hat{s}}}_{2}}.{{{\hat{s}}}_{1}} & {{{\hat{s}}}_{2}}.{{{\hat{p}}}_{t,1}}  \\
   {{{\hat{p}}}_{i,2}}.{{{\hat{s}}}_{1}} & {{{\hat{p}}}_{i,2}}.{{{\hat{p}}}_{t,1}}  \\
\end{matrix} \right]\left[ \begin{matrix}
   {{E}_{t,1_s}}  \\
   {{E}_{t,1_p}}  \\
\end{matrix} \right]=\mathbf{T_{(t,1),(i,2)}}\left[ \begin{matrix}
   {{E}_{t,1_s}}  \\
   {{E}_{t,1_p}}  \\
\end{matrix} \right].
\label{eq:einc2}
\end{equation}
Thus, using the plane equation of $S_2$ ($x\cos\alpha +\frac{h}{2}=y\sin\alpha$), the reflected wave $\vec{E}_{r,2}=[E_{r,2_s}\hat{s}_{2}+E_{r,2_p}\hat{p}_{r,2}]\exp(-j\vec{k}_{r,2}\cdot\vec{r})$ is obtained as follows:
\begin{subequations}\label{eq:er2}
 \begin{equation}
   	\left[ \begin{matrix}
   {{E}_{r,2_s}}  \\
   {{E}_{r,2_p}}  \\
\end{matrix} \right]={{H}_{2}}{{\mathbf{\Gamma }}_{2}}\left[ \begin{matrix}
    {{E}_{i,2_s}}  \\
    {{E}_{i,2_p}}  \\
\end{matrix} \right],
\label{eq:er2_components}
\end{equation}
\begin{equation}
    {{k}_{r,2_z}}={{k}_{t,1_z}},
    \label{eq:er2_kzr2}
\end{equation}
\begin{equation}
  {{k}_{r,2_x}}=-{{k}_{t,1_x}}\cos (2\alpha )+{{k}_{t,1_y}}\sin (2\alpha ),
  \label{eq:er2_kxr2}
\end{equation}
\begin{equation}
    {{k}_{r,2_y}}={{k}_{t,1_x}}\sin (2\alpha )+{{k}_{t,1_y}}\cos (2\alpha ),
    \label{eq:er2_kyr2}
\end{equation}
\end{subequations}\\
where ${{H}_{2}}=\exp \boldsymbol{(}-jh(-{{k}_{t,1_x}}\cos \alpha +{{k}_{t,1_y}}\sin\alpha )\boldsymbol{)}$, and ${{\mathbf{\Gamma }}_{2}}= \begin{bsmallmatrix}
   {{\Gamma }_{2_{s}}} & 0  \\
   0 & {{\Gamma }_{2_{p}}}  \\
\end{bsmallmatrix}$ is the Fresnel reflection coefficient matrix for the incidence at $S_2$ (A complete derivation for Eq. \eqref{eq:er2} and $H_2$ is given in appendix \ref{sec:appendixB}). The dispersion relation for ${{\vec{E}}_{r,2}}$ is given by ${{k}_{r,2_z}}=\sqrt{{{({{n}}k_{in})}^{2}}-k_{r,2_x}^{2}-k_{r,2_y}^{2}}$.

${{\vec{E}}_{t,3}}$ is found by the refraction of ${{\vec{E}}_{r,2}}$ at $S_3$ ($z=-x\cos \alpha +y\sin \alpha +d-\frac{h}{2}$). Again, we rewrite $\vec{E}_{r,2}$ as $\vec{E}_{i,3}=[E_{i,3_s}\hat{s}_{3}+E_{i,3_p}\hat{p}_{i,3}]\exp(-j\vec{k}_{r,2}\cdot\vec{r})$, in which
\begin{equation}
    \left[ \begin{matrix}
   {{E}_{i,3_s}}  \\
   {{E}_{i,3_p}}  \\
\end{matrix} \right]=\left[ \begin{matrix}
   {{{\hat{s}}}_{3}}.{{{\hat{s}}}_{2}} & {{{\hat{s}}}_{3}}.{{{\hat{p}}}_{r,2}}  \\
   {{{\hat{p}}}_{i,3}}.{{{\hat{s}}}_{2}} & {{{\hat{p}}}_{i,3}}.{{{\hat{p}}}_{r,2}}  \\
\end{matrix} \right]\left[ \begin{matrix}
   {{E}_{r,2_s}}  \\
   {{E}_{r,2_p}}  \\
\end{matrix} \right]=\boldsymbol{{T}_{(r,2),(i,3)}}\left[ \begin{matrix}
   {{E}_{r,2_s}}  \\
   {{E}_{r,2_p}}  \\
\end{matrix} \right].
\label{eq:einc3}
\end{equation}
Therefore, the refracted wave $\vec{E}_{t,3}=[E_{t,3_s}\hat{s}_{3}+E_{t,3_p}\hat{p}_{t,3}]\exp(-j\vec{k}_{t,3}\cdot\vec{r})$, which is the final output wave ${\vec{E}}_{F_o}$, is found from the following equations:
\begin{subequations}\label{eq:et3}
 \begin{equation}
   	\left[ \begin{matrix}
   {{E}_{t,3_s}}  \\
   {{E}_{t,3_p}}  \\
\end{matrix} \right]={{{H}_{3}}}\boldsymbol{\tau _{3}}\left[ \begin{matrix}
{{E}_{i,3_s}} \\
    {{E}_{i,3_p}}  \\
\end{matrix} \right],
\label{eq:et3_components}
\end{equation}
\begin{equation}
    {{k}_{t,3_x}}={{k}_{o_x}}={{k}_{r,2_x}}+\cos (\alpha )\left( {{k}_{t,3_z}}-{{k}_{r,2_z}} \right),
    \label{eq:et3_kxt3}
\end{equation}
\begin{equation}
 {{k}_{t,3_y}}={{k}_{o_y}}={{k}_{r,2_y}}+\sin (\alpha )\left( {{k}_{r,2_z}}-{{k}_{t,3_z}} \right),
 \label{eq:et3_kyt3}
\end{equation}
\end{subequations}\\
where ${{H}_{3}}=\exp \boldsymbol{(}-j(d-\frac{h}{2})({{k}_{r,2_z}}-{{k}_{t,3_z}})\boldsymbol{)}$, and $\boldsymbol{\tau _{3}}=\begin{bsmallmatrix}
   {{\tau }_{3_{s}}} & 0  \\
   0 & {{\tau }_{3_{p}}}  \\
\end{bsmallmatrix}$ is the transmission coefficient matrix for the incindence at $S_3$. Note that ${k}_{o_z}={k}_{t,3_z}$.

Eventually, ${{\vec{E}}_{F_o}}=[E_{F_{o_y}}\hat{s}_o+E_{F_{o_{xz}}}{\hat{p}}_{o}]\exp(-j\vec{k}_{o}\cdot\vec{r})$ will be found by converting the polarization of ${{\vec{E}}_{t,3}}$ into the global TE and TM polarizations as follows:
\begin{equation}
 	\left[\begin{matrix}
   {{E}_{F_{o_y}}}  \\
   {{E}_{F_{o_{xz}}}}  \\
\end{matrix} \right]=\left[ \begin{matrix}
   {{{\hat{s}}}_{o}}.{{{\hat{s}}}_{3}} & {{{\hat{s}}}_{o}}.{{{\hat{p}}}_{t,3}}  \\
   {{{\hat{p}}}_{o}}.{{{\hat{s}}}_{3}} & {{{\hat{p}}}_{o}}.{{{\hat{p}}}_{t,3}}  \\
\end{matrix} \right]\left[ \begin{matrix}
   {{E}_{t,3_s}}  \\
   {{E}_{t,3_p}}  \\
\end{matrix} \right]=\boldsymbol{{T}_{(t,3),(o)}}\left[ \begin{matrix}
   {{E}_{t,3_s}}  \\
   {{E}_{t,3_p}}  \\
\end{matrix} \right],
\label{eq:eo_components}
\end{equation}
where $\hat{s}_o=\hat{y}$, and ${{\hat{p}}_{o}}=\frac{\hat{y}\times {{{\vec{k}}}_{o}}}{\left| \hat{y}\times {{{\vec{k}}}_{o}} \right|}$. The dispersion relation ${{k}_{o_z}}=\sqrt{{k_{in}^{2}}-k_{o_x}^{2}-k_{o_y}^{2}}$ is held for ${\vec{E}}_{F_o}$.

By returning from Eq. \eqref{eq:eo_components} to Eq. \eqref{eq:einc1} step by step, and using the dispersion relations of ${{\vec{E}}_{t,1}}$ and ${{\vec{E}}_{F_o}}$, along with the input wave dispersion ${{k}_{in_z}}=\sqrt{{k_{in}^{2}}-k_{in_x}^{2}-k_{in_y}^{2}}$, we can rewrite ${{\vec{E}}_{F_o}}$ and its spatial frequencies in terms of the input Fourier transform and the input spatial frequencies as follows:
\begin{subequations}\label{eq:eo}
    \begin{equation}
\left[\begin{matrix}
   {{E}_{F_{o_y}}}  \\
   {{E}_{F_{o_{xz}}}}  \\
\end{matrix} \right]=\boldsymbol{C}\left[\begin{matrix}
   a_1  \\
   a_2  \\
\end{matrix} \right]F_{in}({{k}_{in_x}},{{k}_{in_y}})H_1H_2H_3\exp \boldsymbol{(}-j({{k}_{o_x}}x+{{k}_{o_y}}y+{{k}_{o_z}}z)\boldsymbol{)}, 
    \end{equation}
    \begin{equation}
        {{k}_{o_x}}=-{{k}_{in_x}}\cos (2\alpha )+{{k}_{in_y}}\sin (2\alpha )
        \label{eq:eo_kxo},
    \end{equation}
    \begin{equation}
        {{k}_{o_y}}={{k}_{in_x}}\sin (2\alpha )+{{k}_{in_y}}\cos (2\alpha ),
        \label{eq:eo_kyo}
    \end{equation}
    \begin{equation}
        {{k}_{o_z}}={{k}_{t,3_z}}={{k}_{in_z}}
        \label{eq:eo_kzo},
    \end{equation}
\end{subequations}\\
where $\mathbf{C}=\begin{bsmallmatrix}
   {C}_{1} & {C}_{2}  \\
   {C}_{3} & {C}_{4}  \\
\end{bsmallmatrix}=\mathbf{{T}_{(t,3),(o)}}\boldsymbol{{{\tau}}_{3}}\mathbf{{T}_{(r,2),(i,3)}}\mathbf{{{\Gamma }}_{2}}\mathbf{{T}_{(t,1),(i,2)}}\boldsymbol{{{\tau }}_{1}}\mathbf{{T}_{(F_{in}),(i,1)}}$ is the coefficient matrix that accounts for cross-polarizations and Fresnel losses (The derivation of Eq. \eqref{eq:eo} is given in appendix \ref{sec:appendixC}), $a_1=\hat{a}_{in}\cdot\hat{s}_{in}$, and $a_2=\hat{a}_{in}\cdot\hat{p}_{in}$.

Note that $E_{F_{o_{xz}}}$ has different polarizations for different wave vectors $\vec{k}_o$. Thus, it is required to decompose $E_{F_{o_{xz}}}$ into separate polarization components $E_{F_{o_x}}=a_3E_{F_{o_{xz}}}$ and $E_{F_{o_z}}=a_4E_{F_{o_{xz}}}$, where $a_3=\hat{p}_{o}\cdot\hat{x}$, and $a_4=\hat{p}_{o}\cdot\hat{z}$. 
Now, by using Eq. \eqref{eq:eo}, the complex amplitude of the output image at $z=d$ can be found by the following integrals:
\begin{subequations}
\begin{equation}
	\begin{split}
	    {{U}_{o_y}}(x,y,d)&=\frac{1}{2\pi}\int\limits_{-\kappa}^{\kappa}{\int\limits_{-\kappa}^{\kappa}{{{E}_{F_{o_y}}}}}d{{k}_{in_x}}d{{k}_{in_y}}=\frac{1}{2\pi}\int\limits_{-\kappa}^{\kappa}\int\limits_{-\kappa}^{\kappa}F_{in}({{k}_{in_x}},{{k}_{in_y}})H\left(C_1a_1+C_2a_2\right)\\&
        \times\exp\left[-j{{k}_{in_x}}(-x\cos (2\alpha)+y\sin (2\alpha))-j{{k}_{in_y}}(x\sin (2\alpha )+y\cos (2\alpha ))\right]\\&\times d{{k}_{in_x}}d{{k}_{in_y}},
    \end{split}
    \label{eq:uyo}
\end{equation}
\begin{equation}
	\begin{split}
{{U}_{o_x}}(x,y,d)&=\frac{1}{2\pi}\int\limits_{-\kappa}^{\kappa}{\int\limits_{-\kappa}^{\kappa}{{{E}_{F_{o_x}}}}}d{{k}_{in_x}}d{{k}_{in_y}}=\frac{1}{2\pi}\int\limits_{-\kappa}^{\kappa}\int\limits_{-\kappa}^{\kappa}F_{in}({{k}_{in_x}},{{k}_{in_y}})H\left(C_3a_1+C_4a_2\right)a_3\\&
        \times\exp\left[-j{{k}_{in_x}}(-x\cos (2\alpha)+y\sin (2\alpha))-j{{k}_{in_y}}(x\sin (2\alpha )+y\cos (2\alpha ))\right]\\&\times d{{k}_{in_x}}d{{k}_{in_y}},
    \end{split}
    \label{eq:uxo}
\end{equation}
\begin{equation}
	\begin{split}
{{U}_{o_z}}(x,y,d)&=\frac{1}{2\pi}\int\limits_{-\kappa}^{\kappa}{\int\limits_{-\kappa}^{\kappa}{{{E}_{F_{o_z}}}}}d{{k}_{in_x}}d{{k}_{in_y}}=\frac{1}{2\pi}\int\limits_{-\kappa}^{\kappa}\int\limits_{-\kappa}^{\kappa}F_{in}({{k}_{in_x}},{{k}_{in_y}})H\left(C_3a_1+C_4a_2\right)a_4\\&
        \times\exp\left[-j{{k}_{in_x}}(-x\cos (2\alpha)+y\sin (2\alpha))-j{{k}_{in_y}}(x\sin (2\alpha )+y\cos (2\alpha ))\right]\\&\times d{{k}_{in_x}}d{{k}_{in_y}},
    \end{split}
    \label{eq:uzo}
\end{equation}
\label{eq:uo}
\end{subequations}\\
where $H=H_1H_2H_3\exp(-j\sqrt{{k_{in}^2}-k_{in_{x}}^2-k_{in_y}^2}d)$, which is called the transfer function, and $\kappa=\frac{2\pi\sin(\frac{\pi}{6})}{\lambda}$. We did not use the range $(-\infty, \infty)$ because the waves with spatial frequencies higher than $k_{in}=\frac{2\pi\sin(\frac{\pi}{2})}{\lambda}$ are non propagating and cannot reach the prism. We also did not use $\frac{\pm2\pi\sin(\frac{\pi}{2})}{\lambda}$ as integral limits because of some considerations that will be discussed in Sec. \ref{sec:dc}. Notice that our derivations assume that the input and output planes are tangent to the leftmost and rightmost edges of the Dove prism, i.e., $z=0$ and $z=d$. If the input or output plane of a given problem is displaced from these assumed positions, we must multiply $H$ by the additional free space transfer functions corresponding to these displacements.

\begin{figure}[!ht]
\includegraphics{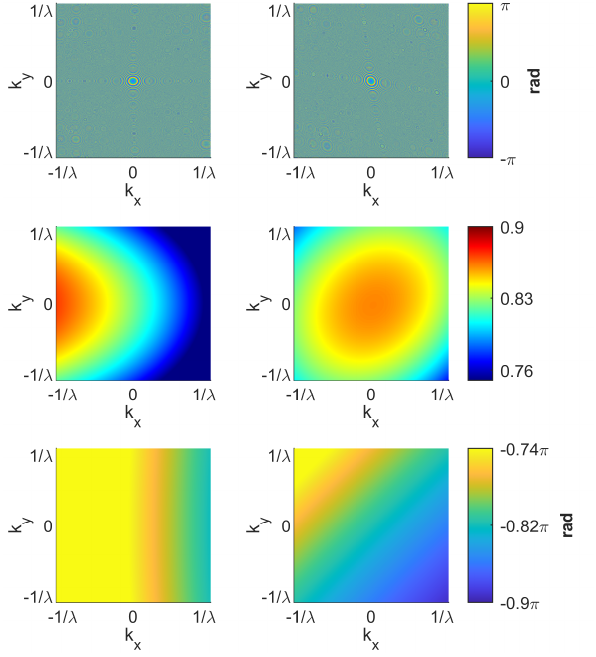}
\caption{\label{fig:HC} The top, middle, and bottom rows illustrate the variation of the complex phase of $H$, $|C_{1}|^2$, and phase of $C_{1}$, respectively, as a function of the input spatial frequencies $k_{in_x}$ and $k_{in_y}$. The left column corresponds to $\alpha=0$, and the right column corresponds to $\alpha=\frac{\pi}{4}$.}
\end{figure}

The matrix of coefficient $\mathbf{C}$, the transfer function $H$, and the coefficients $a_2$, $a_3$, and $a_4$ are, in general, functions of the direction of input propagation. This is because the incident angles $\theta_{i,1}$, $\theta_{i,2}$, and $\theta_{i,3}$, and the polarization unit vectors $\hat{p}_{in}$ and $\hat{p}_o$ are themselves dependent on $k_{in_x}$ and $k_{in_y}$. Figure \ref{fig:HC} shows the variation of the phase of $H$ (top), the squared magnitude of $C_1$ (middle) and the phase of $C_1$ (bottom), when $k_{in_x}$ and $k_{in_y}$ change between $-\frac{1}{\lambda}$ and $\frac{1}{\lambda}$, covering a range greater than the paraxial approximation. It is obvious from the figure that neither $H$ nor $C$ can be considered a constant. In addition, the rotation of the Dove prism changes the distribution of $H$ and $\mathbf{C}$. As an example, the two cases $\alpha=0$ (left) and $\alpha=\frac{\pi}{4}$ (right) are considered in Fig. \ref{fig:HC}, showing a slight tilt in the pattern of $H$ and a significant change in the magnitude and phase of $C_1$. 

\begin{figure}[!ht]
\includegraphics{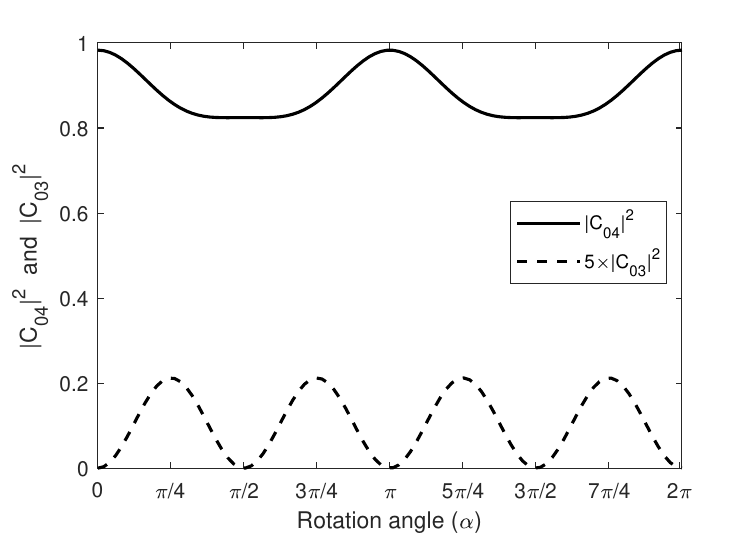}
\caption{\label{fig:C0403} Variation of $|C_{0_4}|^2$ and $|C_{0_3}|^2$ in the paraxial regime as the function of rotation angle $\alpha$.}
\end{figure}

Nevertheless, if the paraxial approximation is kept for the input beam, any element of $\mathbf{C}$ can be treated as a constant, because in the paraxial regime the variation range of input spatial frequencies with respect to the beam direction is small (all Fourier components with considerable optical power lie within a cone with a half angle ${{\theta }_{0}}=\frac{\lambda }{\pi {{W}_{0}}}$, where $W_0$ is the waist radius). Therefore, for small values of $k_{in_x}$ and $k_{in_y}$, $\mathbf{C}({{k}_{in_x}},{{k}_{in_y}})$ can be approximated by the constant matrix $\mathbf{{C}_{0}}=\mathbf{C(0,0)}$ as follows: 
\begin{equation}
	\mathbf{{C}_{0}}=
\left[\begin{matrix}
   {C}_{0_1} & {C}_{0_2}  \\
   {C}_{0_3} & {C}_{0_4}  \\
\end{matrix}\right]
=\left[ \begin{matrix}
   M{{\cos }^{2}}\alpha -N{{\sin }^{2}}\alpha & \frac{1}{2}\left(M+N\right)\sin(2\alpha)  \\
   \frac{1}{2}\left(M+N\right)\sin(2\alpha) & M{{\sin }^{2}}\alpha -N{{\cos }^{2}}\alpha   \\
\end{matrix} \right],
\label{eq:c0}
\end{equation} 
where $M=\tau_{3_{s_0}}\Gamma_{2_{s_0}}\tau_{1_{s_0}}$ and $N=\tau_{3_{p_0}}\Gamma_{2_{p_0}}\tau_{1_{p_0}}$. Here, the Fresnel coefficients correspond to incident angles $\theta_{i,1}=\frac{\pi}{4}$, $\theta_{i,2}=\arcsin(\frac{\sqrt{2}}{2n})+\frac{\pi}{4}$, and $\theta_{i,3}=\arcsin(\frac{\sqrt{2}}{2n})$, respectively, for the indices $1$ to $3$. Again, since $k_{in_x}$ and $k_{in_y}$ are nearly  zero, $\hat{p}_{in}\approx\hat{p}_o\approx\hat{x}$, leading to $a_2=\hat{a}_{in}\cdot\hat{x}$, $a_3=1$, and $a_4=0$. Figure \ref{fig:C0403} displays the squared magnitude of $C_{0_4}$ ($x$ and $z$ combined polarizations) and $C_{0_3}$ (the $y$-polarized input component that couples into the output polarization $x$-$z$) as a function of the rotation angle $\alpha$. The graph is similar to the result given in \cite{padgettDovePrismsPolarized1999}.

In addition to $\mathbf{C}$, the transfer function $H$, which accounts for the propagation effect of the beam, can also be simplified for a paraxial beam. It can be taken out of the integral as the complex constant $H_0=H(0,0)$ given by Eq. \eqref{eq:H0}. In this case, $H_0$ only represents the phase shift between the input and output of a Dove prism, without including the beam divergence.
\begin{equation}
 {{H}_{0}}=\exp \boldsymbol{(}\frac{jk_{in}d}{2}(1+\sqrt{2n^{2}-1}) \boldsymbol{)}.
 \label{eq:H0}
\end{equation}
The exponential argument in Eq. \eqref{eq:uo} mathematically reveals why a Dove prism rotates its input image.  With variable changes ${x}''=-x\cos (2\alpha )+y\sin (2\alpha )$ and ${y}''=x\sin (2\alpha )+y\cos (2\alpha )$, we can rewrite the exponential function in the form of $\exp \boldsymbol{(}-j({{k}_{in_x}}{x}''+{{k}_{in_y}}{y}'')\boldsymbol{)}$. Taking the constant coefficients from the integral and using the new exponential form, we can simplify Eq. \eqref{eq:uo} as follows: 
\begin{subequations}
\begin{equation}
\begin{split}
{{U}_{o_y}}(x,y,d)&=\frac{{{H}_{0}}\left(C_{0_1}a_1+C_{0_2}a_2\right)}{2\pi}\int\limits_{-\kappa }^{\kappa }{\int\limits_{-\kappa }^{\kappa }{F_{in}({{k}_{in_x}},{{k}_{in_y}})\exp\boldsymbol{(}-j({{k}_{in_x}}{x}''+{{k}_{in_y}}{y}'')\boldsymbol{)}}}\,d{{k}_{in_x}}d{{k}_{in_y}}\\&={{H}_{0}}\left(C_{0_1}a_1+C_{0_2}a_2\right){{U_{in}}}({x}'',{y}'',0),
\end{split}
\label{eq:uyo_simplified}
\end{equation}
and similarly,
\begin{equation}
\begin{split}
{{U}_{o_x}}(x,y,d)={{H}_{0}}\left(C_{0_3}a_1+C_{0_4}a_2\right){{U_{in}}}({x}'',{y}'',0).
\end{split}
\label{eq:uxo_simplified}
\end{equation}
\label{eq:uo_simplified}
\end{subequations}\\
Note that $U_{o_z}$ is approximately zero for a paraxial beam ($a_4=0$). In the above equations, we used the variable change
\begin{equation}
	\left[ \begin{matrix}
   y  \\
   x  \\
\end{matrix} \right]=\left[ \begin{matrix}
   \cos (2\alpha ) & -\sin (2\alpha )  \\
   \sin (2\alpha ) & \cos (2\alpha )  \\
\end{matrix} \right]\left[ \begin{matrix}
   1 & 0  \\
   0 & -1  \\
\end{matrix} \right]\left[ \begin{matrix}
   {{y}''}  \\
   {{x}''}  \\
\end{matrix} \right].
\label{eq:rotation}
\end{equation}
In Eq. \eqref{eq:rotation} we rewrote $(x,y)$ in terms of the pair $({x}'',{y}'')$. Leaving aside the coefficients, we realize from Eq. \eqref{eq:uo_simplified} that each point $({x}'',{y}'')$ on the input image transfers invariant to the point $(x,y)$ on the output image after two steps; first, $({x}'',{y}'')$ is mirrored with respect to the horizontal axis (the $y$-axis in this paper). Second, it rotates through $2\alpha$ around the $z$-axis, which is twice the rotation angle of the Dove prism. This formulation gives a comprehensive mathematical proof for the fact that a Dove prism always rotates its input image twice as its rotation angle. Hence, if the paraxial property is held up, the effect of a Dove prism on its input image can be summarized into the following equation:  
\begin{equation}
\left[\begin{matrix}
U_{o_y}\\
U_{o_x}\\
\end{matrix}\right]
=\mathbf{{C}_{0}}{{H}_{0}}
\left[\begin{matrix}
a_1\\
a_2\\
\end{matrix}\right]
Ro{{t}_{2\alpha }}\left[ Mirro{{r}_{h-axis}}({{U_{in}}})\right],
\label{eq:dove_effect}
\end{equation}
where $Rot_{2\alpha}$ and $Mirror_{h-axis}$ represent a $2\alpha$ counterclockwise image rotation and a reflection with respect to the horizontal axis, respectively.

There are two important facts to be noted. First, mirroring always occurs since there is a reflection at $S_2$, even if the Dove prism is not rotated. Second, if the input beam is not paraxial, Eq. \eqref{eq:rotation} is still valid. However, $H$ and $\mathbf{C}$ will be functions of the input spatial frequencies and may distort the output image from a purely rotated version of the input.   

One of the most important applications of a Dove prism is in OAM-based systems, in which the complex amplitude of the beam has a coefficient of $\exp(jl\phi'')$, where $l$ is an integer called the OAM number, and the azimuth angle $\phi''=\arctan(\frac{x''}{y''})$ is the angle between the positive horizontal axis (here the positive $y$-axis) and the line from the origin to the point $(x'',y'')$. Usually, in such applications, a rotation is applied to the OAM-carrying beam by a Dove prism. From Eq. \eqref{eq:dove_effect}, we realize that the azimuth angle $\phi''$ first converts to $\phi'=-\phi''$ because of the mirroring effect, and then to $\phi=\phi'+2\alpha$ because of the rotational effect, where $\phi$ is the azimuth angle corresponding to the output point $(x,y)$. Therefore, the coefficient $\exp(jl\phi)$ on the output plane will actually be $\exp\boldsymbol{(}jl(-\phi''+2\alpha)\boldsymbol{)}=\exp(-jl\phi'')\exp(jl2\alpha)$, explaining, first, why the OAM sign changes and, second, why the term $\exp(jl2\alpha)$ appears for an OAM-carrying beam.  
\section{\label{sec:simulations}Simulations and results}
\begin{figure}[!ht]
\includegraphics{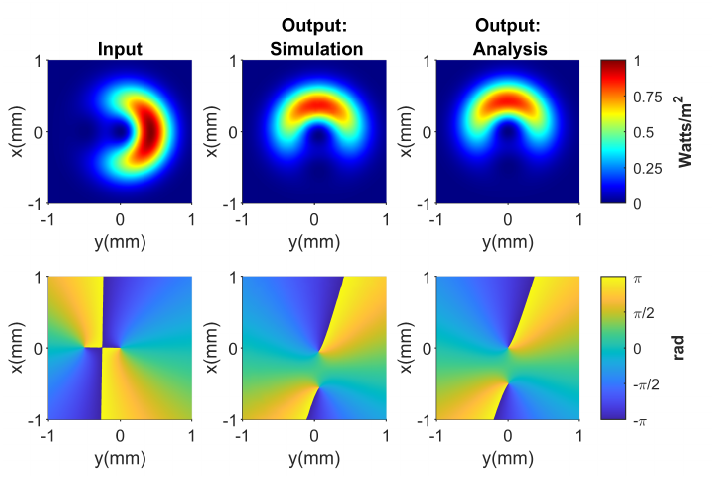}
\caption{\label{fig:simuli}Results of the first simulation. Intensity profiles (upper row) and phase distributions (lower row). The left, middle, and right columns respectively correspond to the input, the output obtained by ray tracing, and the output calculated from Eq. \eqref{eq:uyo_simplified}.}
\end{figure}

In order to validate the derived formulae, two full wave simulations were conducted using the COMSOL Ray Optics module and compared to the results obtained from Eq. \eqref{eq:uyo_simplified}. In the first simulation, a $y$-polarized input beam consisting of two Laguerre-Gaussian modes $LG_{10}$ and $LG_{20}$ \cite{salehFundamentalsPhotonics2019} is considered to impinge on a Dove prism with $\alpha =\frac{\pi }{4}$, $n=1.5$, $d=21.1$ mm, and $h=4.9108$ mm. The results, which are illustrated in Fig. \ref{fig:simuli}, represent that the derived formula Eq. \eqref{eq:uyo_simplified} correctly calculates both the intensity and the phase of the $y$-component of the output electric field.  
As expected, the intensity profile undergoes a $\SI{90}{\degree}$ counter-clockwise rotation. Since the input intensity profile is symmetric with respect to the horizontal axis, the mirroring effect is not observable. However, it can be deduced from the phase patterns, as the color order has been reversed from blue-yellow-green-blue to blue-green-yellow-blue if we sweep counterclockwise. Note that the output phase profiles do not show an exact rotation of $\SI{90}{\degree}$, because there is an additional phase change due to the contribution of $H_0(C_{0_1}a_1+C_{0_2}a_2)$.  

\begin{figure}[!ht]
\includegraphics{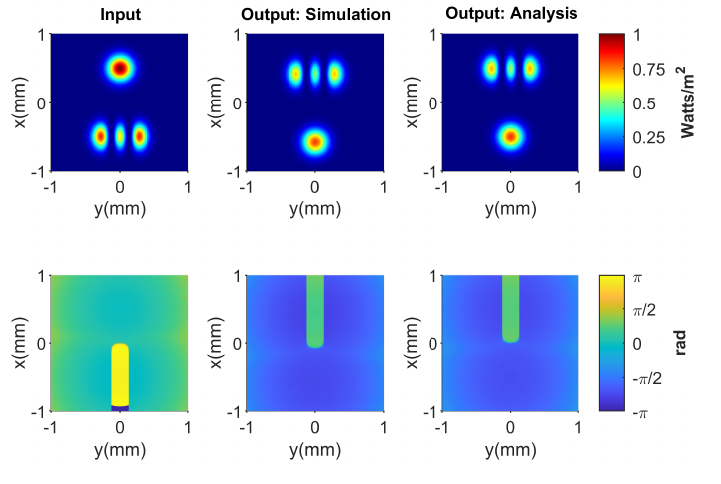}
\caption{\label{fig:simuli_mirror}Results of the second simulation. Intensity profiles (upper row) and phase distributions (lower row). The left, middle, and right columns respectively correspond to the input, output obtained by ray tracing, and output calculated from Eq. \eqref{eq:uyo_simplified}.}
\end{figure}

In order to specifically show the mirroring effect on intensity pattern, we conducted a second simulation in which the input beam is not symmetric with respect to the horizontal axis and the Dove prism rotation angle is zero to exclude the rotational effect. The input we considered is the sum of two Hermite-Gaussian modes $HG_{00}$ and $HG_{02}$ \cite{salehFundamentalsPhotonics2019}. The $HG_{02}$ mode is rotated by $\frac{\pi}{2}$; therefore, it contains three horizontally distributed spots. To make the input beam asymmetric, both modes are laterally shifted from the origin, as shown in Fig. \ref{fig:simuli_mirror}. The mirror effect causes $HG_{00}$ and $HG_{02}$ to exchange their $y$ positions in the intensity pattern.     
The results calculated by Eq. \eqref{eq:uyo_simplified} are well compatible with the ray-tracing results. Again, the output phase is not a precise mirrored image of the input due to $H_0(C_{0_1}a_1+C_{0_2}a_2)$.   

It is worth mentioning that the ray-tracing solver creates rays that emerge from the input spot, all parallel to the beam propagation direction (the $z$-axis). This propagation model fails to account for beam divergence or non-paraxial beams. On the other hand, the COMSOL Wave Optics module directly solves Maxwell's equations over a set of discrete mesh points. Although the results are high-fidelity, meshing a bulk structure like a Dove prism requires millions of points, making this module impractical for simulating the wave-optical output of a Dove prism due to limited memory capacity and computation time. Therefore, the necessity of a known analytical formula for simulating a Dove prism in optical setups is evident. If the problem demands using Eq. \eqref{eq:uo} directly instead of Eq. \eqref{eq:uo_simplified}, the integral has to be numerically calculated; nevertheless, the computation cost will be much less than that of a commercial wave-optical solver. In addition, the numerical solution of the integral of Eq. \eqref{eq:uo} can be made easier and faster by using mathematical tools such as the Fast Fourier Transform (FFT). 
\section{\label{sec:discussion}Scope of the derivation and Additional Remarks}
Our formula is valid and consistent with all physical aspects of the electromagnetic theory. To clarify this further, here we discuss the physical phenomena that the derived formula in Eq. \eqref{eq:uo} encompasses. On the other hand, despite the fact that we performed a wave-optical analysis to find the output of a rotated Dove prism, there are still some minor but worth mentioning considerations that can slightly deviate the result of our given formula from the truly exact solution. In the following, the most important factors are discussed.
\subsection{Physical Phenomena Covered by the Formula}
Because we used the wave nature of light rather than geometrical optics, the phase information of the complex electric field is retained at the output. This enables the calculation of wave interference and consequently the correct prediction of the intensity pattern. In addition, polarization is also incorporated into the derived formula, allowing us to calculate Fresnel losses and cross polarizations. Note that, in contrast to \cite{padgettDovePrismsPolarized1999} and \cite{morenoPolarizationTransformingProperties2003}, we have coupled the polarization and phase contributions within the integral. This is useful in applications where precise calculation of phase shift or cross polarization is needed—for instance, in a quantum processing system that uses the polarization state to store information while simultaneously dealing with highly detailed images (i.e., high spatial frequencies).

Beyond polarization and phase, our derivation also captures non-paraxial effects that are neglected in simpler models. In problems where the beam is highly divergent, a simple paraxial model will not calculate the increase in spot size or the decrease in intensity at the output. This situation is analogous to an input image having high spatial frequencies, which creates image distortion and affects the desired rotational feature in system engineering. Furthermore, unwanted beam shifts may become significant due to the Goos–Hänchen effect for non-paraxial beams. All these effects can be predicted using our derivation, making compensation calculations easier and more accurate.
\subsection{Entrance from the other sides of a Dove prism}
\begin{figure}[!ht]
\includegraphics{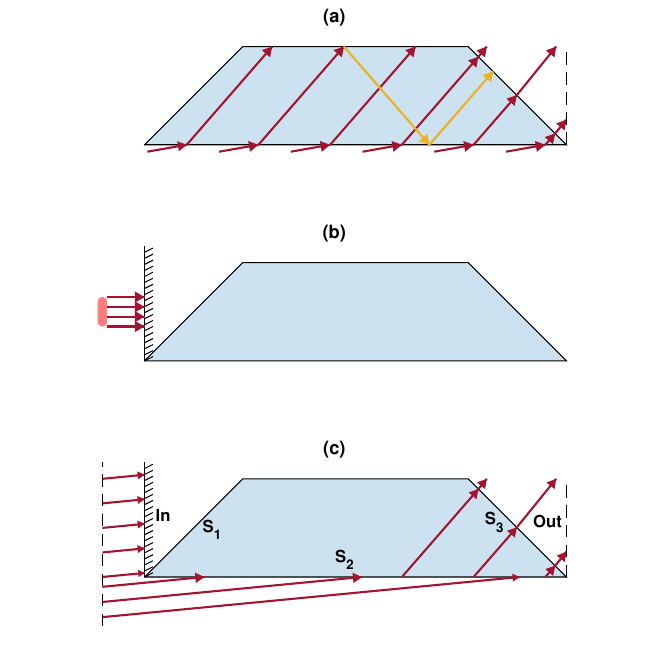}
\caption{\label{fig:discussion} (a) Rays in crimson show a Fourier plane wave that enters from the bottom side of a Dove prism with $\alpha=0$. The wave then refracts from the right side and hits the output plane, which is represented by a dashed line. The orange ray represents a secondary set of rays that can also hit the output after two reflections from the upper and lower sides. (b) The thought experiment. An obstacle (shown as a line with hash marks) blocks the input rays of a spot (red ellipse) from entering the left side of the prism. The obstacle extends from the lowest edge of the prism to positive infinity. (c) The input spot in (b) is replaced with one of its constituting Fourier components. A portion of the component is blocked by the obstacle, while the other portion leaks into the prism from the bottom side. The dashed lines on the left and right show the input and output planes of the prism.}
\end{figure}

We generate a formula for the output of a Dove prism by the assumption that the Fourier components of the input only enter from the left oblique side ($S_1$). This assumption arises from intuition that for an input beam having a spot size much smaller than the prism aperture, the rays only see $S_1$ in their path. However, when the input is decomposed into its constituent plane waves, the intuition vanishes as plane waves extend from minus to positive infinity in free space, producing rays that can hit the lower and upper sides of the prism as well. Figure \ref{fig:discussion}(a) schematically shows this matter with an exemplary case where an upward Fourier plane wave can hit the bottom side of a Dove prism with $\alpha=0$. 
Even if the Fourier plane wave has a small angle with the $z$-axis, it refracts from the bottom side with a steep slope and can reach the output through a refraction from the right oblique side ($S_3$). This unwanted wave from the bottom side also produces secondary waves by its reflections inside the prism, that can find a way to the output. In addition, the downward Fourier components can enter inside the prism and reach the output after reflection from the bottom side, which is not illustrated here for the sake of simplicity. All of this raises the question of whether this wave leakage affects the accuracy or even the correctness of Eq. \eqref{eq:uo}. The answer is that this effect is nonzero and must be attributed to the wave nature of light; nevertheless, it is significantly small, particularly for paraxial cases. The reason is that although Fourier components hit the output plane from the unexpected path, their interference from that path will always be destructive, resulting in a very small intensity in the output. We justified this argument through two different approaches. In the first approach, we directly performed a numerical calculation to find the optical power from the interference of the Fourier plane waves entering from the bottom side (In-$S_2$-$S_3$-Out), then compared the result with the total power of the original spot in the input plane. The calculation showed that, for a Gaussian input beam with a spot size ($W_0$) of about one tenth of the Dove prism aperture ($h$), the fraction of the unexpected power is below $0.2\%$ of the total input, an amount negligible compared to the power from the main path (In-$S_1$-$S_2$-$S_3$-Out). Note that all the Fresnel coefficients for the two refractions are deliberately set to unity to calculate the power leakage under the worst conditions. 

In the second approach, one can conduct a thought experiment in which an obstacle, such as a mirror, is placed between the input spot and the primary side of the prism, blocking the path In-$S_1$-$S_2$-$S_3$-Out. The experiment is schematically illustrated in Fig. \ref{fig:discussion}(b), showing an infinite reflecting plate that starts from the lowermost edge of the prism, cutting the rays of the input spot before they hit the left side. We already know that in a physical setup, the output plane in such an experiment will be dark. On the other hand, the calculated output of Eq. \eqref{eq:uo} is also zero, as is obvious from Fig. \ref{fig:discussion}(c) (Consider that the reflecting plate is equivalent to a step function $u(-x-\frac{h}{2})$ multiplied by the complex amplitude of the input beam at $z=0$. This leads to an exactly zero electric field on the output if Eq. \eqref{eq:dove_effect} is used). The superposition principle implies that the part of the power corresponding to the path In-$S_2$-$S_3$-Out is obtained from the difference between the total power in the output, which is expected to be zero, and the power received from In-$S_1$-$S_2$-$S_3$-Out, which is also zero from the formula. Thus, the output power on the bottom side of the Dove prism must be zero.
\subsection{Total Internal Reflection (TIR) occurrence}
We previously mentioned that each plane wave entering a Dove prism undergoes TIR at $S_2$, but this is not the case for all values of $k_{in_x}$, $k_{in_y}$, and $\alpha$. As an example, for the case $k_{in_x}=\frac{2\pi\sin(30^\circ)}{\lambda}$, $k_{in_y}=\frac{2\pi\sin(33.8182^\circ)}{\lambda}$, $\alpha=21.8182^\circ$, and when the glass refraction index is $n=1.5$, TIR does not occur. Nevertheless, if the angles that the wavevector makes with the $y-z$ and $x-z$ planes are less than $30^\circ$, TIR occurs for every $\alpha$. Note that even if TIR does not occur, the derived formulae are still valid because we treated reflection at $S_2$ in a general manner.
\subsection{\label{sec:dc}Very high spatial frequencies}
It is possible for plane waves with very large $k_{in_x}$ or $k_{in_y}$ not to enter the prism. For example, waves with $k_{in_x}\ge\frac{2\pi\sin\left(45^\circ\right)}{\lambda}$ and $k_{in_y}=0$ cannot refract from $S_1$ when $\alpha=0$. Generally, if the angle between the input wavevector and the normal of $S_1$ pointing into the prism ($\hat{n}_1$ in Fig. \ref{fig:dove}) is greater than $90^\circ$, i.e. $\arccos\left(\frac{{{\vec{k}}_{in}}\cdot{\hat{n}_1}}{\left|{\vec{k}}_{in}\right|}\right)\ge\frac{\pi}{2}$, the plane wave cannot enter the prism and consequently is not involved in the rotation effect. Thus, the limits of the integral in Eq. \eqref{eq:uo} have to be truncated to include only the waves that can enter the prism for a specific rotation angle $\alpha$. Nevertheless, a safe range for both $k_{in_x}$ and $k_{in_y}$ can be determined so that any plane wave with spatial frequencies inside this range is guaranteed to enter the prism for every rotation angle. We find from numerical analysis that the maximum symmetric range is $-\frac{2\pi\sin\left(30^\circ\right)}{\lambda}<k_{in_x}=k_{in_y}<\frac{2\pi\sin\left(30^\circ\right)}{\lambda}$. Therefore, we used this range as the integral limits of  Eq. \eqref{eq:uo}.
\section{\label{sec:conclusion}Conclusion}
Using Fourier optics, we derived a comprehensive closed-form formula for determining the complex amplitude of the electric field at the output plane of a rotated Dove prism. The formula obtained prove the rotational effect of a Dove prism, that is, a rotated Dove prism turns the input image by twice its rotation angle. As two verification examples, we implemented the simplified paraxial version of the main formula to find the output wave for two specific input images incident upon a Dove prism with a \SI{45}{\degree} rotation in one experiment and a zero rotation angle in the second. Comparisons with simulation results obtained using the COMSOL Ray Optics module for the same setup validate our formulae.

Our derivation method has three advantages compared to Dove models in the previous works. First, it includes both polarization and phase information. Second, it is precise and contains no approximations. Third, it is valid for non-paraxial beams, covering a wide range of applications involving high spatial frequencies. This is a significant advantage, as alternative approaches like ray-tracing-based solvers or full numerical Maxwell solvers are either inaccurate or extremely demanding in terms of computation time and capacity for such applications.  
\appendix
\section{\label{sec:appendixA}Fresnel coefficients}
\begin{equation}
    \tau_{s}=\frac{2\eta_t\cos\theta_{i}}{\eta_t\cos\theta_{i}+\eta_{i}\cos\theta_t}
    \label{eq:tauTE}
\end{equation}
\begin{equation}
        \tau_{p}=\frac{2\eta_t\cos\theta_{i}}{\eta_t\cos\theta_{t}+\eta_{i}\cos\theta_{i}}
    \label{eq:tauTM}
\end{equation}
\begin{equation}
    \Gamma_{s}=\frac{\eta_t\cos\theta_{i}-\eta_{i}\cos\theta_t}{\eta_t\cos\theta_{i}+\eta_{i}\cos\theta_t}
    \label{eq:gammaTE}
\end{equation}
\begin{equation}
        \Gamma_{p}=\frac{\eta_t\cos\theta_{t}-\eta_{i}\cos\theta_{i}}{\eta_t\cos\theta_{t}+\eta_{i}\cos\theta_{i}}
    \label{eq:gammaTM}
\end{equation}
\section{\label{sec:appendixB}Derivation of $H_2$}
First, we make some modifications to the plane equation of $S_2$ by multiplying it separately by $\cos \alpha $ and $\sin \alpha $ as follows:
\begin{equation} 
\begin{split}
  &x\cos \alpha +\frac{h}{2}=y\sin\alpha \Rightarrow x{{\cos }^{2}}\alpha +\frac{h}{2}\cos \alpha =y\sin \alpha \cos \alpha  \\ 
 & \Rightarrow x\left( \frac{1+\cos 2\alpha }{2} \right)+\frac{h}{2}\cos \alpha =\frac{y}{2}\sin 2\alpha  \\ 
 & \Rightarrow x=y\sin 2\alpha -x\cos 2\alpha -h\cos \alpha, 
\end{split}
\label{eq:S2eq_modified1}
\end{equation}
and
\begin{equation}   
\begin{split}
  & x\cos \alpha +\frac{h}{2}=y\sin \alpha \Rightarrow x\cos \alpha \sin \alpha +\frac{h}{2}\sin \alpha =y{{\sin }^{2}}\alpha  \\ 
 & \Rightarrow \frac{x}{2}\sin 2\alpha +\frac{h}{2}\sin \alpha =y-y\left( \frac{1+\cos 2\alpha }{2} \right) \\ 
 & \Rightarrow y=x\sin 2\alpha +y\cos 2\alpha +h\sin \alpha. 
\end{split}
\label{eq:S2eq_modified2}
\end{equation}
By applying the continuity equations of the electric and magnetic tangential components for TE polarization, we can form the following equations based on the general forms given in Eq. \eqref{eq:tangentialte_eqs}:
\begin{equation}
\begin{split}
 &{{E}_{i,2_s}}\exp \boldsymbol{(}-j({{k}_{t,1_x}}x+{{k}_{t,1_y}}y+{{k}_{t,1_z}}z)\boldsymbol{)}+{{E}_{r,2_s}}\exp \boldsymbol{(}-j({{k}_{r,2_x}}x+{{k}_{r,2_y}}y+{{k}_{r,2_z}}z)\boldsymbol{)}\\&={{E}_{t,2_s}}\exp \boldsymbol{(}-j({{k}_{t,2_x}}x+{{k}_{t,2_y}}y+{{k}_{t,2_z}}z)\boldsymbol{)},   
\end{split}
\label{eq:s2_tangte1}
\end{equation}
and
\begin{equation}
    \begin{split}
      & \frac{-\cos ({{\theta }_{i,2}}){{E}_{i,2_s}}}{{{\eta }_{i,2}}}\exp \boldsymbol{(}-j({{k}_{t,1_x}}x+{{k}_{t,1_y}}y+{{k}_{t,1_z}}z)\boldsymbol{)}\\&+\frac{\cos ({{\theta }_{r,2}}){{E}_{r,2_s}}}{{{\eta }_{i,2}}}\exp \boldsymbol{(}-j({{k}_{r,2_x}}x+{{k}_{r,2_y}}y+{{k}_{r,2_z}}z)\boldsymbol{)}\\&=\frac{-\cos ({{\theta }_{t,2}}){{E}_{t,2_s}}}{{{\eta }_{t,2}}}\exp \boldsymbol{(}-j({{k}_{t,2_x}}x+{{k}_{t,2_y}}y+{{k}_{t,2_z}}z)\boldsymbol{)}.
    \end{split}
    \label{eq:s2_tangte2}
\end{equation}	
Note that $\theta_{i,2}=\theta_{r,2}$ from the law of reflection. By replacing $x$ and $y$ in the exponential terms of $\vec{E}_{i,2_s}$ in Eq. \eqref{eq:s2_tangte1} and Eq. \eqref{eq:s2_tangte2} respectively with the results from Eq. \eqref{eq:S2eq_modified1} and Eq. \eqref{eq:S2eq_modified2}, and rearranging, we will get:
\begin{equation}
  \begin{split}
      & {{E}_{i,2_s}}\exp\boldsymbol{(} -jx\left( {{k}_{t,1_y}}\sin 2\alpha -{{k}_{t,1_x}}\cos 2\alpha  \right)\boldsymbol{)}\times \exp \boldsymbol{(} -jy\left( {{k}_{t,1_x}}\sin 2\alpha +{{k}_{t,1_y}}\cos 2\alpha  \right) \boldsymbol{)}\\&\times \exp \left( -j{{k}_{t,1_z}}z \right)\times \exp \boldsymbol{(} -jh\left( {{k}_{t,1_y}}\sin \alpha -{{k}_{t,1_x}}\cos \alpha  \right) \boldsymbol{)}+{{E}_{r,2_s}}\exp \left( -j{{k}_{r,2_x}}x \right)\\&\times \exp \left( -j{{k}_{r,2_y}}y \right)\times\exp \left( -j{{k}_{r,2_z}}z \right)\\&={{E}_{t,2_s}}\exp \left( -j{{k}_{t,2_x}}x \right)\times \exp \left( -j{{k}_{t,2_y}}y \right)\times \exp \left( -j{{k}_{t,2_z}}z \right),
  \end{split}
  \label{eq:s2_tangte1_modified}
\end{equation}
and
\begin{equation}
    \begin{split}
       & \frac{-\cos ({{\theta }_{i,2}}){{E}_{i,2_s}}}{{{\eta }_{i,2}}}\exp \boldsymbol{(} -jx\left( {{k}_{t,1_y}}\sin 2\alpha -{{k}_{t,1_x}}\cos 2\alpha  \right) \boldsymbol{)}\times\\& \exp \boldsymbol{(} -jy\left( {{k}_{t,1_x}}\sin 2\alpha +{{k}_{t,1_y}}\cos 2\alpha  \right) \boldsymbol{)}\times \exp \left( -j{{k}_{t,1_z}}z \right)\\&\times \exp \boldsymbol{(} -jh\left( {{k}_{t,1_y}}\sin \alpha -{{k}_{t,1_x}}\cos \alpha  \right) \boldsymbol{)}+\frac{\cos ({{\theta }_{r,2}}){{E}_{r,2_s}}}{{{\eta }_{i,2}}}\exp \left( -j{{k}_{r,2_x}}x \right)\\&\times \exp \left( -j{{k}_{r,2_y}}y \right)\times \exp \left( -j{{k}_{r,2_z}}z \right)\\&=\frac{-\cos ({{\theta }_{t,2}}){{E}_{t,2_s}}}{{{\eta }_{t,2}}}\exp \left( -j{{k}_{t,2_x}}x \right)\times \exp \left( -j{{k}_{t,2_y}}y \right)\times \exp \left( -j{{k}_{t,2_z}}z \right). 
    \end{split}
    \label{eq:s2_tangte2_modified}
\end{equation}
Since the above equations must be valid for all values of $x$, $y$, and $z$ on $S_2$, the exponential terms that are functions of these variables must be united. This will give us Eq. \eqref{eq:er2_kzr2} to Eq. \eqref{eq:er2_kyr2}. We then eliminate these exponential functions from both sides of Eq. \eqref{eq:s2_tangte1_modified} and Eq. \eqref{eq:s2_tangte2_modified}. Thus, they reduce to:
\begin{equation}
  {{E}_{i,2_s}}\exp \boldsymbol{(} -jh\left( {{k}_{t,1_y}}\sin \alpha -{{k}_{t,1_x}}\cos \alpha  \right) \boldsymbol{)}+{{E}_{r,2_s}}={{E}_{t,2_s}},
  \label{eq:s2_tangte1_simplified}
\end{equation}
and
\begin{equation}
\begin{split}
    &\frac{-\cos ({{\theta }_{i,2}}){{E}_{i,2_s}}}{{{\eta }_{i,2}}}\exp \boldsymbol{(} -jh\left( {{k}_{t,1_y}}\sin \alpha -{{k}_{t,1_x}}\cos \alpha  \right) \boldsymbol{)}+\frac{\cos ({{\theta }_{r,2}}){{E}_{r,2_s}}}{{{\eta }_{i,2}}}\\&=\frac{-\cos ({{\theta }_{t,2}}){{E}_{t,2_s}}}{{{\eta }_{t,2}}}.
    \label{eq:s2_tangte2_simplified}
\end{split}
\end{equation}
From Eq. \eqref{eq:s2_tangte1_simplified} and Eq. \eqref{eq:s2_tangte2_simplified}, we can find the reflected wave ${{E}_{r,2_s}}$ in terms of ${{E}_{i,2_s}}$ as follows:
\begin{equation}
\begin{split}
  {{E}_{r,2_s}}&=\exp \boldsymbol{(} -jh\left( {{k}_{t,1_y}}\sin \alpha -{{k}_{t,1_x}}\cos \alpha  \right) \boldsymbol{)}\frac{{{\eta }_{t,2}}\cos ({{\theta }_{i,2}})-{{\eta }_{i,2}}\cos ({{\theta }_{t,2}})}{{{\eta }_{t,2}}\cos ({{\theta }_{i,2}})+{{\eta }_{i,2}}\cos ({{\theta }_{t,2}})}{{E}_{i,2_s}}\\&={{H}_{2}}{{\Gamma }_{ga_{te}}}{{E}_{i,2_s}}. 
\end{split} 
\end{equation}
The same result is obtained for $H_2$ in the case of TM polarization.
\section{\label{sec:appendixC}Derivation of the output spatial frequencies}
First, using the dispersion relation for ${{k}_{t,1_z}}$ and the equations \eqref{eq:et1_kxt1} and \eqref{eq:et1_kyt1}, we can find ${{k}_{t,1_z}}$ as follows:
\begin{equation}
 n^{2}{k_{in}^{2}}=k_{t,1_x}^{2}+k_{t,1_y}^{2}+k_{t,1_z}^{2}\Rightarrow {{k}_{t,1_z}}=\frac{{{k}_{in_z}}-A+\sqrt{{{(A+{{k}_{in_z}})}^{2}}-2{k_{in}^{2}}(1-n^{2})}}{2},
 \label{eq:kzt1_A}
\end{equation}
where $A=-{{k}_{in_x}}\cos (\alpha )+{{k}_{in_y}}\sin (\alpha )$. Note that we must take the positive square root when solving the quadratic equation for $k_{t,1_z}$, since $k_{t,1_z}$ must be positive for every $\alpha$ to make a forward-propagating wave.
Similarly, applying the dispersion relation for ${{\vec{E}}_{o}}$ and using Eq. \eqref{eq:et3_kxt3} and Eq. \eqref{eq:et3_kyt3}, we have:
\begin{equation}
{k_{in}^{2}}=k_{o_x}^{2}+k_{o_y}^{2}+k_{o_z}^{2}\Rightarrow {{k}_{o_z}}=\frac{{{k}_{r,2_z}}+{A}'+\sqrt{{{({{k}_{r,2_z}}-{A}')}^{2}}-2{k_{in}^{2}}(n^{2}-1)}}{2},
\label{eq:kzo_Aprim}
\end{equation}
where ${A}'=-{{k}_{r,2_x}}\cos (\alpha )+{{k}_{r,2_y}}\sin (\alpha )$. Again, we must choose the positive root for $k_{o_z}$. Substituting Eq. \eqref{eq:er2_kxr2} and Eq. \eqref{eq:er2_kyr2} into $A'$, we can rewrite $A'$ as follows:
\begin{equation}
 \begin{split}
  {A}'&={{k}_{t,1_x}}\cos (\alpha )\cos (2\alpha )-{{k}_{t,1_y}}\sin (2\alpha )\cos (\alpha )+{{k}_{t,1_x}}\sin (2\alpha )\sin (\alpha )+{{k}_{t,1_y}}\sin (\alpha )\cos (2\alpha ) \\ 
 & ={{k}_{t,1_x}}\cos (\alpha )-{{k}_{t,1_y}}\sin (\alpha ). \\ 
\end{split}   
\end{equation}
Then, using Eq. \eqref{eq:et1_kxt1} and Eq. \eqref{eq:et1_kyt1}, we obtain the following:
\begin{equation}
 \begin{split}
  {A}'&=\left[ \cos (\alpha )({{k}_{in_z}}-{{k}_{t,1_z}})+{{k}_{in_x}} \right]\cos (\alpha )-\left[ \sin (\alpha )({{k}_{t,1_z}}-{{k}_{in_z}})+{{k}_{in_y}} \right]\sin (\alpha )\\&={{k}_{in_z}}-{{k}_{t,1_z}}+{{k}_{in_x}}\cos (\alpha )-{{k}_{in_y}}\sin (\alpha ) 
 ={{k}_{in_z}}-{{k}_{t,1_z}}-A.
\end{split}
\label{eq:Aprim_A}
\end{equation}
Substituting \eqref{eq:kzt1_A} into \eqref{eq:Aprim_A}, we find:
\begin{equation}
\begin{split}
A'&=k_{in_z}-A-\frac{{{k}_{in_z}}-A+\sqrt{{{(A+{{k}_{in_z}})}^{2}}-2{k_{in}^{2}}(1-n^{2})}}{2}
\\&=\frac{{{k}_{in_z}}-A-\sqrt{{{(A+{{k}_{in_z}})}^{2}}-2{k_{in}^{2}}(1-n^{2})}}{2}.
\end{split}
\end{equation}
Thus,
\begin{subequations}
\begin{equation}
k_{t,1_z}-A'=\sqrt{{{(A+{{k}_{in_z}})}^{2}}-2{k_{in}^{2}}(1-n^{2})},
\end{equation}
and
\begin{equation}
k_{t,1_z}+A'=k_{in_z}-A.
\end{equation}
\label{eq:Aktz1_pair}
\end{subequations}\\
Using the results of Eq. \eqref{eq:Aktz1_pair} in Eq. \eqref{eq:kzo_Aprim} and knowing that $k_{r,2_z}=k_{t,1_z}$, we will get the following,
\begin{equation}
    {{k}_{o_z}}=\frac{1}{2}\left[ {{k}_{t,1_z}}+{{k}_{in_z}}-{{k}_{t,1_z}}-A+\sqrt{{{({{k}_{in_z}}+A)}^{2}}} \right]={{k}_{in_z}},
    \label{eq:kzo_is_kz}
\end{equation}
which is the proof for Eq. \eqref{eq:eo_kzo}.

Putting Eq. \eqref{eq:et1_kxt1} and Eq. \eqref{eq:et1_kyt1} in Eq. \eqref{eq:er2_kxr2} and Eq. \eqref{eq:er2_kyr2}, we have:
\begin{equation}
\begin{split}
  {{k}_{r,2_x}}&=-\cos (2\alpha )\cos (\alpha )\left[ {{k}_{in_z}}-{{k}_{t,1_z}} \right]-{{k}_{in_x}}\cos (2\alpha )-\sin (2\alpha )\sin (\alpha )\left[ {{k}_{in_z}}-{{k}_{t,1_z}} \right]\\&+{{k}_{in_y}}\sin (2\alpha )
 =({{k}_{t,1_z}}-{{k}_{in_z}})\cos (\alpha )-{{k}_{in_x}}\cos (2\alpha )+{{k}_{in_y}}\sin (2\alpha),
 \label{eq:kxr2}
\end{split}
\end{equation}
and
\begin{equation}
\begin{split}
 {{k}_{r,2_y}}&=\sin (2\alpha )\cos (\alpha )\left[ {{k}_{in_z}}-{{k}_{t,1_z}} \right]+{{k}_{in_x}}\sin (2\alpha )-\cos (2\alpha )\sin (\alpha )\left[ {{k}_{in_z}}-{{k}_{t,1_z}} \right]\\&+{{k}_{in_y}}\cos (2\alpha )
 =({{k}_{in_z}}-{{k}_{t,1_z}})\sin (\alpha )+{{k}_{in_x}}\sin (2\alpha )+{{k}_{in_y}}\cos (2\alpha ).
 \label{eq:kyr2}
\end{split}
\end{equation}
Now that we have found $k_{r,2_x}$ and $k_{r,2_y}$ in terms of the input spatial frequencies $k_{in_x}$, $k_{in_y}$, and $k_{in_z}$, we can extract Eq. \eqref{eq:eo_kxo} and Eq. \eqref{eq:eo_kyo} by substituting Eq. \eqref{eq:kxr2} and Eq. \eqref{eq:kyr2} into Eq. \eqref{eq:et3_kxt3} and Eq. \eqref{eq:et3_kyt3}, recalling that ${{k}_{t,1_z}}={{k}_{r,2_z}}$ From Eq. \eqref{eq:er2_kzr2}.

\bibliography{Dove_paper}

@article{armitageRotaryShearingInterferometry1965,
  title = {Rotary {{Shearing Interferometry}}},
  author = {Armitage, J.D and Lohmann, A},
  year = {1965},
  month = apr,
  journal = {Optica Acta: International Journal of Optics},
  volume = {12},
  number = {2},
  pages = {185--192},
  publisher = {Taylor \& Francis},
  issn = {0030-3909},
  doi = {10.1080/713817933}
}

@article{bolducHighresolutionSurfacePlasmon2009,
  title = {High-Resolution Surface Plasmon Resonance Sensors Based on a Dove Prism},
  author = {Bolduc, Olivier R. and Live, Ludovic S. and Masson, Jean-Fran{\c c}ois},
  year = {2009},
  month = mar,
  journal = {Talanta},
  volume = {77},
  number = {5},
  pages = {1680--1687},
  issn = {0039-9140},
  doi = {10.1016/j.talanta.2008.10.006}
}

@book{chengFieldWaveElectromagnetics1989,
  title = {Field and {{Wave Electromagnetics}}},
  author = {Cheng, David},
  year = {1989},
  edition = {2},
  publisher = {Addison-Wesley},
  address = {Reading, Mass},
  isbn = {978-0-201-12819-2},
}

@article{gonzalezHowDovePrism2006,
  title = {How a {{Dove}} Prism Transforms the Orbital Angular Momentum of a Light Beam},
  author = {Gonz{\'a}lez, N. and {Molina-Terriza}, G. and Torres, J. P.},
  year = {2006},
  month = oct,
  journal = {Optics Express},
  volume = {14},
  number = {20},
  pages = {9093--9102},
  publisher = {Optica Publishing Group},
  issn = {1094-4087},
  doi = {10.1364/OE.14.009093},
  urldate = {2025-08-01},
  copyright = {{\copyright} 2006 Optical Society of America},
  langid = {english}
}

@article{karanQuantifyingPolarizationChanges2022,
  title = {Quantifying Polarization Changes Induced by Rotating {{Dove}} Prisms {{andK-mirrors}}},
  author = {Karan, Suman and {Ruchi} and Mohta, Pranay and Jha, Anand K.},
  year = {2022},
  month = oct,
  journal = {Applied Optics},
  volume = {61},
  number = {28},
  pages = {8302--8307},
  publisher = {Optica Publishing Group},
  doi = {10.1364/AO.472543}
}

@article{leachInterferometricMethodsMeasure2004,
  title = {Interferometric {{Methods}} to {{Measure Orbital}} and {{Spin}}, or the {{Total Angular Momentum}} of a {{Single Photon}}},
  author = {Leach, Jonathan and Courtial, Johannes and Skeldon, Kenneth and Barnett, Stephen M. and {Franke-Arnold}, Sonja and Padgett, Miles J.},
  year = {2004},
  month = jan,
  journal = {Physical Review Letters},
  volume = {92},
  number = {1},
  pages = {013601},
  publisher = {American Physical Society},
  doi = {10.1103/PhysRevLett.92.013601},
  urldate = {2025-12-18}
}

@article{morenoDovePrismIncreased2003,
  title = {Dove Prism with Increased Throughput for Implementation in a Rotational-Shearing Interferometer},
  author = {Moreno, Ivan and Paez, Gonzalo and Strojnik, Marija},
  year = {2003},
  month = aug,
  journal = {Applied Optics},
  volume = {42},
  number = {22},
  pages = {4514--4521},
  publisher = {Optica Publishing Group},
  issn = {2155-3165},
  doi = {10.1364/AO.42.004514},
  urldate = {2025-08-10},
  copyright = {{\copyright} 2003 Optical Society of America},
  langid = {english}
}

@article{morenoJonesMatrixImagerotation2004,
  title = {Jones Matrix for Image-Rotation Prisms},
  author = {Moreno, Ivan},
  year = {2004},
  month = jun,
  journal = {Applied Optics},
  volume = {43},
  number = {17},
  pages = {3373--3381},
  issn = {1559-128X},
  doi = {10.1364/ao.43.003373},
  langid = {english}
}

@article{morenoPolarizationTransformingProperties2003,
  title = {Polarization Transforming Properties of {{Dove}} Prisms},
  author = {Moreno, Ivan and Paez, Gonzalo and Strojnik, Marija},
  year = {2003},
  month = may,
  journal = {Optics Communications},
  volume = {220},
  number = {4},
  pages = {257--268},
  issn = {0030-4018},
  doi = {10.1016/S0030-4018(03)01423-8}
}

@article{padgettDovePrismsPolarized1999,
  title = {Dove Prisms and Polarized Light},
  author = {Padgett, Miles J. and Lesso, J. Paul},
  year = {1999},
  month = feb,
  journal = {Journal of Modern Optics},
  volume = {46},
  number = {2},
  pages = {175--179},
  publisher = {Taylor \& Francis},
  issn = {0950-0340},
  doi = {10.1080/09500349908231263},
  urldate = {2025-08-01}
}

@book{salehFundamentalsPhotonics2019,
  title = {Fundamentals of {{Photonics}}},
  author = {Saleh, Bahaa and Teich, Malvin},
  year = {2019},
  month = feb,
  edition = {3rd},
  publisher = {Wiley},
  isbn = {978-1-119-50687-4},
  langid = {english}
}

@article{sar-elRevisedDovePrism1991,
  title = {Revised {{Dove}} Prism Formulas},
  author = {{Sar-El}, H. Z.},
  year = {1991},
  month = feb,
  journal = {Applied Optics},
  volume = {30},
  number = {4},
  pages = {375--376},
  publisher = {Optica Publishing Group},
  issn = {2155-3165},
  doi = {10.1364/AO.30.000375},
  urldate = {2025-08-10},
  copyright = {{\copyright} 1991 Optical Society of America},
  langid = {english}
}

@book{smithModernOpticalEngineering2008,
  title = {Modern {{Optical Engineering}}},
  author = {Smith, Warren J.},
  year = {2008},
  edition = {4},
  publisher = {McGraw Hill},
  address = {New York},
  isbn = {978-0-07-147687-4},
  langid = {english}
}

@article{wangSinglepathSagnacInterferometer2017,
  title = {Single-Path {{Sagnac}} Interferometer with {{Dove}} Prism for Orbital-Angular-Momentum Photon Manipulation},
  author = {Wang, Fang-Xiang and Chen, Wei and Li, Ya-Ping and Zhang, Guo-Wei and Yin, Zhen-Qiang and Wang, Shuang and Guo, Guang-Can and Han, Zheng-Fu},
  year = {2017},
  month = oct,
  journal = {Optics Express},
  volume = {25},
  number = {21},
  pages = {24946--24959},
  publisher = {Optica Publishing Group},
  issn = {1094-4087},
  doi = {10.1364/OE.25.024946},
  urldate = {2025-08-01},
  copyright = {{\copyright} 2017 Optical Society of America}
}

@article{versmold_interferometric_2025,
	title = {Interferometric {Amplification} and {Suppression} of {External} {Beam} {Shifts}},
	volume = {135},
	issn = {0031-9007, 1079-7114},
	url = {https://link.aps.org/doi/10.1103/fggq-yhz8},
	doi = {10.1103/fggq-yhz8},
	number = {25},
	urldate = {2026-05-23},
	journal = {Physical Review Letters},
	author = {Versmold, Carlotta and Dziewior, Jan and Huber, Florian and Köster, Elina and Reznik, Gregory and Vaidman, Lev and Weinfurter, Harald},
	month = dec,
	year = {2025},
	pages = {253802},
}

@article{montes-flores_rotationally_2024,
	title = {Rotationally shearing interferometer for exoplanet detection: mathematical derivation, theory, and simulation},
	volume = {32},
	issn = {1094-4087},
	shorttitle = {Rotationally shearing interferometer for exoplanet detection},
	url = {https://opg.optica.org/abstract.cfm?URI=oe-32-27-48391},
	doi = {10.1364/OE.542197},
	number = {27},
	urldate = {2026-05-23},
	journal = {Optics Express},
	author = {Montes-Flores, Manuel and Garcia-Torales, Guillermo and Strojnik, Marija},
	month = dec,
	year = {2024},
	pages = {48391},
}

@article{zhang_hong-ou-mandel_2016,
	title = {Hong-{Ou}-{Mandel} interference of entangled {Hermite}-{Gauss} modes},
	volume = {94},
	url = {https://link.aps.org/doi/10.1103/PhysRevA.94.033855},
	doi = {10.1103/PhysRevA.94.033855},
	number = {3},
	journal = {Phys. Rev. A},
	publisher = {American Physical Society},
	author = {Zhang, Yingwen and Prabhakar, Shashi and Rosales-Guzmán, Carmelo and Roux, Filippus S. and Karimi, Ebrahim and Forbes, Andrew},
	month = sep,
	year = {2016},
	pages = {033855},
}

@article{yao_loss-tolerant_2024,
	title = {Loss-tolerant and supersensitive angular rotation estimation based on quantum-enhanced interferometers},
	volume = {110},
	url = {https://link.aps.org/doi/10.1103/PhysRevA.110.032429},
	doi = {10.1103/PhysRevA.110.032429},
	number = {3},
	journal = {Phys. Rev. A},
	publisher = {American Physical Society},
	author = {Yao, Wenxiu and Zhang, Xiaoli and Tian, Long and Liu, Xuan and Shi, Shaoping and Zheng, Yaohui},
	month = sep,
	year = {2024},
	pages = {032429},
}

@article{lowry_reflection_2021,
	title = {Reflection terahertz time-domain spectroscopy for imaging and identifying concealed interfaces in insulated systems},
	author = {Lowry, Seth N. and Price, Brad D. and Hartley, Ian D. and Shegelski, Mark R. A. and Reid, Matt},
	year = {2021},
	month = aug,
	journal = {Applied Optics},

	volume = {60},
	number = {23},
	pages = {6818--6828},
	issn = {2155-3165},
	doi = {10.1364/AO.429888},
	urldate = {2026-05-23},
}

@article{chen_precise_2017,
	title = {Precise transverse alignment of spatial light modulator sections for complex optical field generation},
	author = {Chen, Jian and Wan, Chenhao and Kong, Lingjiang and Zhan, Qiwen},
	year = {2017},
	month = apr,
	journal = {Applied Optics},

	volume = {56},
	number = {10},
	pages = {2614--2620},
	issn = {2155-3165},
	doi = {10.1364/AO.56.002614},
	urldate = {2026-05-23},
}

@article{xiao_orbital_2018,
	title = {Orbital angular momentum-enhanced measurement of rotation vibration using a {Sagnac} interferometer},
	author = {Xiao, Shixiong and Zhang, Lidan and Wei, Dan and Liu, Fang and Zhang, Yong and Xiao, Min},
	year = {2018},
	month = jan,
	journal = {Optics Express},

	volume = {26},
	number = {2},
	pages = {1997--2005},
	issn = {1094-4087},
	doi = {10.1364/OE.26.001997},
	urldate = {2026-05-20},
	publisher = {Optica Publishing Group},
}

@article{byersSuperresolutionUpgradeDeep2025,
	title = {Super-resolution upgrade for deep tissue imaging featuring simple implementation},
	author = {Byers, Patrick and Kellerer, Thomas and Li, Miaomiao and Chen, Zhifen and Huser, Thomas and Hellerer, Thomas},
	year = {2025},
	month = jun,
	journal = {Nature Communications},
	volume = {16},
	number = {1},
	pages = {5386},
	issn = {2041-1723},
	doi = {10.1038/s41467-025-60744-y},
	urldate = {2026-05-20},
}

\end{document}